\documentclass[twocolumn,showpacs,preprintnumbers,amsmath,amssymb, superscriptaddress]{revtex4-1}

\usepackage{graphicx}
\usepackage{bm}

\begin{document}

\title{The Landau-Ginzburg-Wilson Hamiltonian for the Griffiths Phase}

\author{X. T. Wu}
\affiliation{Department of Physics, Beijing Normal University,
Beijing, 100875, China}

\date{\today}

\begin{abstract}
The Landau-Ginzburg-Wilson Hamiltonian with random temperature for the phase transition in disordered systems from the Griffith phase to ferromagnetic phase is reexamined. From the saddle point solutions, especially the excited state solutions, it is shown that the system self-organizes into blocks coupled with their neighbors like superspins, which are emergent variables. Taking the fluctuation around these saddle point solutions into account, we get an effective Hamiltonian, including the emergent superspins of the blocks, the fluctuation around the saddle point solutions, and their couplings. Applying Stratonovich-Hubbard transformation to the part of superspins, we get a Landau-Ginzburg-Wilson Hamiltonian for the blocks. From the saddle point equations for the blocks, we can get the second generation blocks, of which sizes are much larger than the first generation blocks. Repeating this procedure again and again, we get many generations of blocks to describe the asymptotic behavior. If a field is applied, the effective field on the superspins is multiplied greatly and proportional to the block size. For a very small field, the effective field on the higher generation superspins can be so strong to cause the superspins polarized radically. This can explain the extra large critical isotherm exponent discovered in the experiments. The phase space of reduced temperature vs. field is divided into many layers , in which different generation blocks dominate the critical behavior. The sizes of the different generation emergent blocks are new relevant length scales.  This can explain a lot of puzzles in the experiments and the Monte Carlo simulation.
\end{abstract}


\maketitle

\section{Introduction}
There are a large amount of theoretical and experimental works on the phase transition and critical phenomena in the disordered systems. However the understanding on this topic is far from satisfactory. For the pure system, the Landau-Ginzburg-Wilson (LGW) Hamiltonian is very successful to understand the phase transition and critical phenomena \cite{wilson,fisher}. On the contrary, the counterpart for the disordered system,  the LGW Hamiltonian with random temperature is poorly understood. Not mention to explain the experiments consistently.

The greatest challenge comes from the experiments on the Griffiths phase, especially the experiments for the doped MnO materials \cite{salamon,salamon1,triki}. It is very hard to understand the data found in these experiments within the existing framework of theory. According to the Harris criterion \cite{harris} and the renormalization group (RG) theory \cite{harrisa,khm,grinstein}, the disorder is irrelevant or irrelevant depending on the specific heat exponent of the pure systems. The critical behaviors of these systems should belong to several universalities.  For some of these materials, the exponents are so anomalous that they can not be cataloged into the known universalities.  Dotsenko argued \cite{dotsenko1999} that nonanalytic nature of the Griffiths phase makes it difficult to apply off-the-shelf RG techniques because that the essential contributions of local minima destroy the length-scaling picture of a random-fixed point universality class. Chan, Goldenfeld and Salamon obtained new scaling relations to explain the experimental data from a heuristic calculation of the leading order essential singularity in the distribution of Yang-Lee zeroes. \cite{salamon1}.

In this paper, I try to show that the phase transition in disordered system is still able to be understood in the framework of LGW Hamiltonian. Our starting point is the LGW Hamiltonian with random temperature. It is reexamined based on the saddle point (SP) solutions \cite{wu1,wu2a,wu2,wu3,wu4}.  Besides the usual ground state solutions, in which the order parameters are along the same direction, there exist a large amount of excited state solutions \cite{wu2}. These excited state solutions are local minima and beyond the usual perturbation technique. In these excited state SP solutions the system self-organizes into blocks coupled with their neighbors like superspins, which are emergent variables to describe the collective motions.

The physical picture of the phase transition from the Griffiths phase to the ordered phase is like follows \cite{wu2}. As the temperature decreased, the SP solution begins to be nonzero in some regions. These regions are locally ordered and are the incipient blocks. In one block the order parameters point to the same direction and the directions of order parameter in different blocks can be different. Even all regions are locally ordered, the long range order is not definitely realized because of the excited state solutions, in which the adjoined blocks interacts with each other like superspins. The superspins are the unit vectors of order parameter's direction of the blocks. Only when the interactions between the blocks become strong enough, the long range order takes place. This model is called block model.

In this paper, the block model is discussed further. Taking the fluctuation around the SP solutions into account, an effective Hamiltonian is obtained. There are three parts in the effective Hamiltonian. The first part is for the superspins of the blocks, the second one is for the fluctuation, and the third one is for the coupling between the previous two parts. These three parts are all relevant in the RG. As the interactions of the superspins become strong enough, the block model become critical and can be described by a new LGW Hamiltonian also. Solving the SP equation for the new LGW Hamiltonian of the superspins, we can get the second generation blocks. The blocks obtained from the SP solution of the original LGW can be called the first generation blocks. This procedure can be repeated again and again, many generations of blocks can be obtained to describe the asymptotic behavior near the critical point. Therefore the phase space of temperature vs. field is divided into many layers, in which each generation blocks dominate the critical behavior. The structure of phase space is like a Chinese (nested) boxes. This can explain a lot of properties discovered in the experiments.

This paper is arranged as follows. In the section II, the LGW Hamiltonian with random temperature is introduced. In section III, the block model for Ising-type and the multi-component system is discussed separately. The excited state SP solutions without external field are reviewed. In section IV the effect of the external field is analysed. In section V, the effective Hamiltonian including the fluctuation around the SP solutions is obtained. In section VI, the Chinese boxes structure in the phase space of temperature vs. field is analysed. Some discussions are given in section VII. It is shown that the block model can explain the anomalous scaling behaviours found in the experiments on Griffiths phase.

\section{The LGW model with random temperature}

We consider the LGW Hamiltonian of with a random temperature
\begin{equation}
H_{LGW} =  \frac{1}{a^d}\int d{\textbf x} \{ \frac{a^2}{2}|\bigtriangledown {\bf m}|^2
   +\frac{t}{2}{\bf m}^2 +{g \over 4}{\bf m}^4 -{\textbf h}\cdot {\bf m} \},
\label{eq:h0}
\end{equation}
where $a$ is the microscopic short distance cutoff (for example, the lattice spacing); the order parameter have $p$ components
\begin{eqnarray}
{\bf m} = ( \phi_1,\phi_2,\cdots,\phi_p), &
\hskip 0.2cm  {\bf m}^2=\sum_{i=1}^{p}m_i^2, \nonumber \\
 |\bigtriangledown {\bf m}|^2  =  \sum_{i=1}^{p}|\bigtriangledown m_i|^2,  & \hskip 0.4cm {\textbf h}\cdot {\bf m} =\sum_{i=1}^{p}h_i m_i
;
\end{eqnarray}
$t=t({\textbf x})=\bar{t}+\tilde{t}({\textbf x})$, and $\bar{t},\tilde{t}({\textbf x})$  are the average reduced temperature and the random part caused by the disorder respectively. We consider the short-range correlated disorder, such as that in diluted  or random bond disorder. The correlation range of the disorder in these cases are just the lattice spacing $a$. As we know, $p=1$ is the Ising-type system; $p=2$ is XY-type and $p=3$ is Heisenburg-type.

The disorder distribution is not of essence. The variance of the disorder is the characterized quantity.
\begin{equation}
\Delta=\sqrt{<\tilde{t}^2({\textbf x})>}.
\end{equation}
where $ <\cdots> $ is the disorder configuration average. Because $|\tilde{t}({\textbf x})|<1$, it is obvious that $\Delta<1$. For the diluted systems with void probability $p$, the variance is given by $\Delta=\sqrt{p(1-p)}$. For random bond systems, the variance of the disorder is proportional to the variance of the bonds.

This LGW Hamiltonian with random temperature is the starting point to study the critical phenomena in disordered systems   \cite{harrisa,khm,grinstein}. More than 40 years ago, using replica trick to average the disorder, a translation-invariant Hamiltonian, has been obtained. In these theories, the saddle point solution is assumed to be zero above the critical temperature. In recent 20 years, this assumption is questioned. In the works of Dotsenko and his collaborators \cite{dotsenko1995,dotsenko1999,tarjus,dotsenko2006,dotsenko2006a}, it is argued that in the disorder dominated region one finds  a macroscopic number of local minimum saddle point equations.  Dotsenko et. al proposed that taking into account other local minimum configurations of the random Hamiltonian may cause the replica symmetry breaking \cite{dotsenko1995}. The regions where the saddle point solutions are not zero are called locally ordered regions (LOR). It should be pointed out that in these theories, that the interaction between the LORs is ignored.

The existence of LOR has been supported by many experiments.  In recent years, LOR is discovered in the disordered magnetic  systems \cite{grig1,grig2,grig3}. The experiments on the superfluid transition of $^4He$ in porous media also revealed the localized Bose condensation above the global superfluid transition temperature \cite{plan1,plan2}. In order to explain recent scanning tunnel microscope STM experiments, an inhomogeneous gapped superconductor with superconducting islands and metallic regions is proposed \cite{galitski}. The existence of Ferromagnetic region in the Paramagnetic phase is discovered \cite{eremina}. Spontaneous magnetization above $T_C$ in polycrystalline LaCaMnO and LaBaMnO is discovered \cite{turcaud}

To understand the LOR, we have studied SP solutions for the LGW Hamiltonian with random temperature in detail in some cases \cite{wu1,wu2a,wu2,wu3,wu4}. In the SP solutions, the physical picture is like follows. As the temperature lowers, there begin to appear some LORs. These LOR are incipient blocks. As the temperature decreases further, more and more LORs appear, and some are adjoined. In the ground state saddle point solution, the adjoined blocks have parameters with the same direction. However, there exist excited SP solutions in which the order parameters in the adjoined blocks have different directions. These excited state solutions are local minima and can not be dealt by the usual perturbation technique properly.

\section{The block model}

\subsection{The Ising case}

To be clear, we discuss the Ising case, i.e. one-component case, and multi-component cases separately. For the Ising case without the external field
the saddle point equation are given by
\begin{equation}
-a^2\bigtriangledown^2 \b{$m$} +[t({\textbf x})
   + g\b{$m$}^2 ]\b{$m$}=0,
\label{eq:spe1}
\end{equation}
Here we denote the SP solution by \b{$m$} rather than $\bar{m}$, which is usually used to denote the average magnetization in mean field theory for the pure system. There are excited state SP solutions, so $\b{$m$}$ are not the average magnetization even on the mean field level. The SP solutions in one and two dimension are studied numerically in detail \cite{wu1,wu2a,wu2}.  There are two ground state solutions, in which it has $\b{m}\ge 0$ or $\b{m}\le 0$ over the whole system. However this is not all. It is shown there exist excited state solutions with domain walls. In these excited state solutions, the system is self-organized into blocks and in each block the order parameter can be positive or negative \cite{wu2}.

It is shown that the concerned temperature range is $ |t|\sim \Delta^{-\frac{4}{4-d}}$.
For $t<\Delta^{-\frac{4}{4-d}}$, the LORs become dense;  for $t<-\Delta^{-\frac{4}{4-d}}$, the interactions between blocks are already strong enough to realize the long range order.

The shapes and sizes of the blocks depend on the disorder configuration. A method called``Opening windows" is proposed to find these elementary blocks \cite{wu3}. Elementary block means that there exists no such excited solution that a domain wall crosses over the block and separate the block into two pieces. In any excited state solution, the domain wall only locates at the periphery of the blocks.  For the Gaussian distribution random temprature the average size of blocks is given by
\begin{equation}
a_I\propto a\Delta ^{-2/(4-d)}
\end{equation}
as shown in reference \cite{wu2}. This indicates that the block size is much larger than the lattice spacing. The subscript $I$ of $a_I$ means the first generation, which will be explained in detail later. The scaling of the SP solutions are discussed in 1 and 2-dimension and the scaling relations agree with the numerical solutions very well. It can be expected that those scaling relations are valid in 3-dimension also. In the following, the estimation for the 3-dimensional systems is based on these scaling relations.

For the Ising case, the $ \nu$th excited state solution can be approximated by
\begin{equation}
\b{$m$}^{(\nu)}\approx \sum_i \psi_i({\bf x}){\bf s}_i^{(\nu)}
\label{eq:sds}
\end{equation}
where
\begin{equation}
\psi_i({\bf x})=\{ \begin{array}{ll}
             \Phi_0({\bf x});   & {\bf x}\in ith ~~ block \\
             0 ;   & other ~~cases.
            \end{array}
\label{eq:excited}
\end{equation}
$\Phi_0({\bf x}) \geq 0$ is the ground state solution. Throughout this paper we use $\Phi_0({\bf x})$ to denote the ground state solution.
${\bf s}_i^{(\nu)}=\pm 1$ is the superspin to denote the direction of the order parameter in the $i$th block. This variable is an emergent and collective axis for the block.

We can draw an analogy between the solutions and landform. For the ground state (the positive solution) the magnitude of the order parameter is like the height.  The excited states are such that some peaks are upside down and below the horizon. The places where are just at the horizon are the domain wall.  Domain walls are usually located at the valleys of the ground state so that the difference between the order parameter value of the ground state and the excited state is small. Then the free energy increase is small.

Substituting the saddle point solution into Eq. (\ref{eq:h0}), one get the free energy \cite{dotsenko1995}
\begin{equation}
F_{\nu}=H(\{\b{$m$}^{(\nu)}\})=-\int d{\textbf x} {g \over 4}
(\b{$m$}^{(\nu)})^4, \label{eq:fe}
\end{equation}
for the $\nu$th solution. Neglecting the fluctuation around the saddle point solutions, the partition function is given by
$ Z_{SP} =\sum_{\nu}e^{-F_{\nu}},$
where $e^{-F_{\nu}}$ is the thermodynamic probability of the
$\nu$th solution.

Considering two adjoined blocks, the order parameters in the two blocks have the same sign in the ground state solution, and opposite sign in the excited state solution. The domain wall between two blocks with opposite order parameter will cause an increase in the free energy. We take this increase due to the domain wall as the coupling between the two blocks. Then we get a Hamiltonian of Ising model with random couplings for the blocks,
\begin{equation}
F_{\nu}\approx F_0
-\sum_{<ij>}K^I_{ij}({\bf s}^I_i \cdot {\bf s}^I_j-1)/2 \label{eq:fren}
\end{equation}
where $F_{\nu}, F_0$ are the free energy of $\nu$th state and ground state, the summation is over nearest neighbors and $K^I_{ij}$ is the contribution due to the segment of domain wall between the blocks $i$ and $j$. The superscript $I$ in each variable is referred to the first generation block. This approximate free energy for the superspins of the blocks is obtained from the numerical calculations \cite{wu2,wu3}.

\subsection{The multi-component case}

For the multi-component cases, the LGW Hamiltonian can also be written by
\begin{eqnarray}
H  =  \frac{1}{a^d}\int d{\textbf x} \{ \frac{a^2}{2} (|\bigtriangledown m|^2+m^2|\bigtriangledown {\textbf n}|^2)   \nonumber \\
    +{t \over 2}m^2 +{g \over 4} m^4 -m{\textbf h}\cdot {\textbf n}  \}£¬\nonumber \\
\label{eq:h-multi}
\end{eqnarray}
where   $ m=\sqrt{\sum_{i=1}^{i=p}{\bf m}_i^2}, ~~\textbf{n}={\bf m}/m$,  $m$ is the magnitude of order parameter and ${\bf n}$ is the unit direction vector of the order parameter. The ground state saddle point solution $\Phi_0$ is given by the Eq. (\ref{eq:spe1}) with $\b{$m$}>0$ and ${\bf n}({\bf x})$ being the same over the whole system.

The ground state SP solution in the multi-component case is the same as that in the Ising case, but the excited state SP solutions are different. In the excited state SP solutions for the Ising-type case, there are domain walls. For the multi-component case, the order parameter rotates its direction by slow transverse fluctuations.

For the excited state SP solutions for the multi-component case, the rotation of the order parameter can be reached by slow transverse fluctuations.
If we consider the direction of the parameter varies slowly in space, then the longitudinal component satisfies
\begin{equation}
a^2(-\bigtriangledown^2+|\bigtriangledown {\textbf n}|^2)\b{m}+(t+g\b{m}^2)\b{m}=0
\end{equation}
with the absence of external field. Because we are concerned with the long wavelength fluctuations, $|\bigtriangledown {\textbf n}|^2$ should be very small and it can be dealt as a perturbation. We set $\b{m}=\Phi_0+\delta \b{m}$, we expand the above equation into
\begin{equation}
-a^2\bigtriangledown^2\Phi_0+(t+g\Phi_0^2)\Phi_0=0,
\end{equation}
and
\begin{equation}
a^2(-\bigtriangledown^2\delta \b{m}+|\bigtriangledown {\textbf n}|^2\Phi_0)+(t+3g\Phi_0^2)\delta \b{m}=0.
\end{equation}
From these two equations, we can get
\begin{equation}
\int 2g\Phi_0^3\delta \b{m} d {\textbf x}=-\int \Phi_0^2|\bigtriangledown {\textbf n}|^2 d {\textbf x}
\end{equation}
The corresponding free energy is given by
\begin{eqnarray}
F & = & -\frac{g}{4}\int\b{m}_h^4 d{\bf x} \nonumber \\
 & = & -\frac{g}{4}\int \Phi_0^4 d{\bf x}+\frac{1}{2}\int \Phi_0^2|\bigtriangledown {\textbf n}|^2 d {\textbf x}
\label{eq:dg}
\end{eqnarray}

The slow transverse fluctuations  can turn the direction of the order parameter to the opposite direction with a very small free energy increase. These solutions are counterparts of the excited states with domain walls for the Ising case. We consider a local region with size much bigger than a block and set the ground state SP solution be $ \b{\bf m}({\bf x})=\Phi_0({\bf x}){\bf e}_1$.
The direction ${\bf e}_1$ is longitudinal, the other $p-1$ directions ${\bf e}_2,\cdots,{\bf e}_p$ are transverse ones. In order to investigate the eigenmodes of the transverse fluctuations, we set
\begin{equation}
 {\bf m}=(\Phi_0+\tilde{m}_1){\bf e}_1+\sum_{i=2}^p \tilde{m}_i{\bf e}_i,
\label{eq:gold}
\end{equation}
where $\tilde{m}_i, i=2, \cdots, p$ are the transverse fluctuation part.
Then the free energy increase due to the transverse fluctuation in Eq. (\ref{eq:dg}) satisfies
\begin{eqnarray}
\delta F_{G} & = & \frac{1}{2}\int \Phi_0^2|\bigtriangledown {\textbf n}|^2 d {\textbf x} \nonumber \\
           & \approx & \frac{1}{a^d}\int d{\textbf x} \sum_{i=2}^p [\frac{a^2}{2} |\bigtriangledown^2 \tilde{m}_i|^2+\frac{t+\Phi_0^2}{2}\tilde{m}_i^2 ].
\label{eq:gaussian-appro1}
\end{eqnarray}
The second equation is proved in the appendix. It can be seen that it is a Gaussian approximation. The eigenmodes of $\tilde{m}_i$ satisfy the following equation
\begin{equation}
-a^2\bigtriangledown^2 \tilde{m}_i+(t+\Phi_0^2)\tilde{m}_i=\lambda \tilde{m}_i.
\end{equation}
For the pure system where $t$ and $\Phi_0$ are homogeneous, the eigenmodes of the above equation are the spin waves. For the disordered systems, $t$ and $\Phi_0$ are not homogeneous, however the equation can be solved numerically. In reference \cite{wu4}, we have studied the eigenmodes numerically for the XY-type case, where it has $p=2$. It is shown that the lowest excited solutions are such that the phases in each block are the same approximately and the phases in different blocks are different. The corresponding excited solutions are given by Eq. (\ref{eq:sds}) and (\ref{eq:excited}), however ${\bf s}_i^{(\nu)}$ is a 2-dimensional unit vector, which is in fact the direction of the order parameter in the $i$th block. It is an emergent and collective axis for the block. This is to say that in the lowest excited states the phases of the order parameter in the ith block can be described by only one variable ${\bf s}_i$. This conclusion can be extended to the Heisenburg case, where $p=3$, and larger component number cases, because the eigenmodes equation of the transverse components are the same. Hence $p$-component systems should have the same Hamiltonian Eq. (\ref{eq:fren}), in which ${\bf s}_i,{\bf s}_j$ are $p$-dimensional unit vectors.

In this paper, we do not discuss the percolative phase transition, in which $g \rightarrow 0$. As shown in \cite{wu2}, $J_{ij}\sim g^{-1}$. In the limit $g \rightarrow 0$, the couplings between neighbored blocks are infinitely large. The adjoined blocks form a cluster and their order parameter point to the same direction.  We are concerned the case with not very small $g$. For example, for Ising model $g=1/3$.
Usually for finite $g$ even the blocks percolate through out the whole system, the long range order does not take place. Only when the interactions between the blocks become strong enough, the long range order takes place.
For $\bar{t}>0$, the coupling between blocks $J_{ij}$ are not zero, but very small. For $\bar{t}<0$, they increase fast, and approximately are given by $J_{ij}\sim g^{-1} a_I^{d-1} |\bar{t}|^{3/2}$.

\section{The effect of external field}

The effect of external field on the SP¡¡solutions is dramatic. The effective field on the superspins is multiplied greatly.

For the Ising case, the  SP equation with an external field is given by
\begin{equation}
-\bigtriangledown^2 \b{$m$}_h +[t+ g\b{$m$}_h^2 ]\b{$m$}_h-h =0.
\label{eq:sad-ising}
\end{equation}
Here we add a subscript in $\b{m}_h$ to tell it from the saddle point solution $\b{${ m}$}$ without the field ${ h}$. If the field is very weak, it can be dealt as a perturbation. Let $\b{ m}_h=\b{m}+\delta \b{m}$, where  $\b{m}$ is given by the saddle point equation without field, Eq. (\ref{eq:spe1}) and the solutions are described by Eq. (\ref{eq:sds}). Expanding $\b{$m$}_h$ in Eq. (\ref{eq:sad-ising}), we get the equation for $\delta \b{$m$}$
\begin{equation}
-\bigtriangledown^2 \delta \b{$m$} +[t+3 g\b{$m$}^2 ]\delta \b{$m$}-h =0.
\end{equation}
From the above two equations, we get
\begin{equation}
\int  2g\b{$m$}^3\delta \b{m}d{\bf x}=\int   h\b{m} d {\bf x}=\int   {\bf h}\cdot \b{\bf m} d {\bf x}
\end{equation}
The corresponding free energy is given by
\begin{eqnarray}
F & = & -\frac{g}{4}\int\b{$m$}_h^4 d{\bf x} \nonumber \\
 & = & -\frac{g}{4}\int\b{$m$}^4 d{\bf x}-\frac{1}{2}\int {\bf h}\cdot \b{\bf m} d{\bf x}
\end{eqnarray}
Considering Eq. (\ref{eq:fren}), at the SP level we get
\begin{equation}
F\approx F_0 -\sum_{<ij>}K^I_{ij}({\bf s}^I_i \cdot {\bf s}^I_j-1)/2 -\sum_i {\bf h}^I_i \cdot {\bf s}_i^I
\label{eq:heff-ex}
\end{equation}
where
\begin{equation}
{\bf h}^I_i=\int_{{\bf x}\in ith~~block}\frac{1}{2}\Phi_0{\bf h}({\bf x})d{\bf x}
\label{eq:feff}
\end{equation}
The effective field on the superspin ${\bf s}_i$ is ${\bf h}^I_i$, which is proportional to the volume of the block. It has $h^I_i\sim h a_I^d \bar{\Phi}_0$. The average of $\Phi_0$ is appoximately $\sim \Delta^{2/(4-d)}$ \cite{wu2}. Then we have $h_i^I \sim h\Delta^{-\frac{2(d-1)}{4-d}}$.  Its physical significance is simple to understand. The effective field on the superspin is proportional to the sum of field on the spins in the block. Usually it has $a_I \sim 10^2 -10^3$ in 3-dimensional system assuming a strong disorder $\Delta=0.3$ according to the scaling \cite{wu2}. Then the effective field is multiplied $10^4-10^6$ times.

If the effective field $h^I_i$ is large enough, say $h^I_i \approx 10$, the field dominates. Then the superspins are almost polarized, only the ground state SP solution accounts. According to the estimate, only if the corresponding fild satisfies $h >10^{-3}-10^{-5}$, the first generation superspins are totally polarized in 3-dimensional system with $\Delta=0.1\sim 0.3$.

For the multi-component case, the SP equation with the external field for the longitudinal component $m$ is given by
\begin{equation}
a^2(-\bigtriangledown^2+|\bigtriangledown {\textbf n}|^2)\b{m}+(t+g\b{m}^2)\b{m}-{\bf h}\cdot {\bf n}=0
\end{equation}
For a very weak field, it can be dealt as perturbation. Through similar derivation in the last subsection, we can get the modification of free energy due to the field. Together with the modification from the transverse fluctuation term $|\bigtriangledown {\textbf n}|^2$, we obtain the same equation as Eq. (\ref{eq:heff-ex}). The effective field on the superspins of block is also given by Eq. (\ref{eq:feff}). However the variables ${\bf s}_i$ in these two equations are replaced by $p$-dimensional unit vectors.

\section{The effective Hamiltonian including the fluctuation around the SP solutions}

The variables ${\bf s}_i$ are emergent collective coordinates. On the SP level, the system is described by these coordinates. To be beyond the SP level, one should take the fluctuation over the saddle point solutions into account.

Considering the fluctuation around the SP solutions, we expand the order parameter as
\begin{equation}
{\bf m}({\bf x})=\b{{\bf m}}^{(\nu)}({\bf x})+\tilde{{\bf m}}({\bf x}),
\end{equation}
where $\tilde{\bf m}$ is the fluctuation around the SP solutions.
Substituting it into GLW Hamiltonian Eq.(\ref{eq:h0}), one gets
\begin{eqnarray}
H & = & \sum_{\nu} \{ F_{\nu}+ \frac{1}{a^d}\int d{\textbf x} \{ \frac{a^2}{2}|\bigtriangledown \tilde{\bf m}|^2
   +\frac{t'}{2}\tilde{\bf m}^2 +{g \over 4}\tilde{\bf m}^4  \} \nonumber \\
& & +\int d{\bf x} g[(\b{{\bf m}}^{(\nu)}\cdot \tilde{{\bf m}})\tilde{m}^2+\frac{1}{2}(\b{{\bf m}}^{(\nu)}\cdot \tilde{{\bf m}})^2] \},
\label{eq:heff}
\end{eqnarray}
where $t'({\bf x})=t({\bf x})+g\Phi_0^2({\bf x})$. The linear part for the fluctuation $\tilde{m}$ is zero, so the filed term is absent. The field term is absorbed in the $F_{\nu}$. Substituting Eq. (\ref{eq:sds}) and (\ref{eq:fren}) into the above equation, on obtains the effective Hamiltonian
\begin{eqnarray}
H_I & = & -\frac{1}{2}\sum_{<i,j>} K^I_{ij}{\bf s}^I_i \cdot {\bf s}^I_j -\sum_i {\bf h}^I_i \cdot {\bf s}^I_i \nonumber \\
   & & +  \frac{1}{a^d}\int d{\textbf x} \{ \frac{a^2}{2}|\bigtriangledown \tilde{\bf m}|^2
   +\frac{t}{2}\tilde{\bf m}^2 +{g \over 4}\tilde{\bf m}^4  \} \nonumber \\
  & & +\frac{g}{a^d}\int d{\textbf x} [{\Phi}^{0}({\bf s}^I\cdot \tilde{\bf m})\tilde{m}^2+\frac{1}{2}\Phi_0^2({\bf s}^I\cdot \tilde{\bf m})^2] .
\label{eq:heff-1}
\end{eqnarray}
where the constant terms in Eq. (\ref{eq:fren})  are ignored,
$ {\bf s}^I({\bf x})= {\bf s}^I_i$ for ${\bf x}\in $ ith block. The superscript $I$ in $K_{ij}^I$ and $S_i^I$ is referred to the first generation blocks. In this approximation, the partition function is given by
$ Z=\int Ds \int D \tilde{{\bf m}} \exp (-H_I)$,
where it has  $\int Ds=\sum_{\{s_i\}}$ for Ising-type and  $\int D{\bf s}=\prod_i \int d{\bf s}_i$ for other cases.
Now we have two sets of order parameters. The first one is for the superspins of blocks, and the second one is for the fluctuation around the SP.

As the $J_{ij}^I$ become large, the Hamiltonian for the superspins become critical. Then the correlation length for the superspins become much larger than the block size $a_I$. We can apply the Stratonovich-Hubbard transformation to the random bond Hamiltonian for the superspins, then we get a LGW Hamiltonian for the blocks and the critical fluctuations:
\begin{eqnarray}
H_I & = & \frac{1}{a_I^d}\int d{\textbf x} \{ \frac{a_I^2}{2}|\bigtriangledown {\bf m}^I|^2
          +\frac{t^I}{2}({\bf m}^I)^2 +{g^I \over 4}({\bf m}^I)^4 \nonumber \\
  &&   -{\bf h}^I \cdot {\bf m}^I \}+\frac{g}{a^d} \int d{\textbf x} \Phi_0 \tilde{m}^2\tilde{{\bf m}}\cdot {\bf m}^I \nonumber \\
    & &+ \frac{1}{a^d}\int d{\textbf x} \{ \frac{a^2}{2}|\bigtriangledown \tilde{\bf m}|^2
   +\frac{t'}{2}\tilde{\bf m}^2 +{g \over 4}\tilde{\bf m}^4  \}
\label{eq:hI}
\end{eqnarray}
where $ {\bf m}^I$ is the variable corresponding to the block spin ${\bf s}_i^I$. The terms $({\bf m}^I)^2,({\bf m}^I)^4,({\bf m}^I\cdot \tilde{{\bf m}})\tilde{m}^2$ are kept in the transformation and the higher order terms are neglected. The last term in Eq. (\ref{eq:heff-1}) is absorbed in the redefinition of $t'$  for the fluctuation $\tilde{m}$ in the transformation. Since we are interested only in the long wave length modes of ${\bf m}^I({\textbf x})$, we also use the approximation of space continuation.  The reduced temperatures $t^I({\textbf x})$ are determined by the random couplings $K^I_{ij}$ between the blocks. The last term is the coupling between the blocks and the critical fluctuations.  In the transformation, we only keep the random temperature $t^I({\bf x})$ to represent the effect of spatial fluctuation of $K^I_{ij}$. Other terms fluctuating in space are ignored because these terms are irrelevant in RG.

Similarly we can write $t^I({\bf x})=\bar{t}^I+\tilde{t}^I({\bf x})$, where the variance of $\tilde{t}^I({\bf x})$ is the ratio between the variance the averge of $(K^I)^{-1}_{ij}$. We denote it $\Delta_I=\sqrt{<\tilde{t}_I({\bf x})^2>}$.

As $\bar{t}^I<\Delta_I^{4/(4-d}$, the SP equation for $m_I({\textbf x})$ becomes nonzero, and we should also get excited state solutions. Then on the saddle point level, one will get second generation blocks and a block model again. The block size of the second generation should be
\begin{equation}
a_{II}\propto a_I\Delta_I^{-\frac{2}{4-d}}
\end{equation}
following Eq. (6).

Let $m^I=\b{m}^I+\tilde{m}^I$, where $\b{m}^I$ is the SP solution. One can write the effective Hamiltonian with the presence of second generation blocks:
\begin{eqnarray}
H_{II} & = & -\frac{1}{2}\sum_{<i,j>} K^{II}_{ij}{\bf s}^{II}_i \cdot {\bf s}^{II}_j -\sum_i {\bf h}^{II}_i \cdot {\bf s}^{II}_i \nonumber \\
& & + \frac{1}{a_I^d}\int d{\textbf x} [ \frac{a_I^2}{2}|\bigtriangledown \tilde{\bf m}^I|^2
   +\frac{t'^I}{2}(\tilde{\bf m}^I)^2 +{g^I \over 4}(\tilde{\bf m}^I)^4 ]\nonumber \\
& &+ \frac{1}{a^d}\int d{\textbf x} [ \frac{a^2}{2}|\bigtriangledown \tilde{\bf m}|^2
   +\frac{t'}{2}\tilde{\bf m}^2 +{g \over 4}\tilde{\bf m}^4]  \nonumber \\
& & +\frac{g^I}{a_I^d}\int d{\textbf x}[({\Phi}^I_0({\bf s}^{II}\cdot \tilde{\bf m} ^I) (\tilde{\bf m} ^I)^2
+\frac{1}{2}(\Phi^I_0)^2({\bf s}^{II}\cdot \tilde{\bf m}^I)^2] \nonumber \\
& &  +\frac{g}{a^d} \int d{\textbf x} \Phi_0 [\tilde{m}^2 \tilde{{\bf m}}\cdot \tilde{\bf m}^I]  .
\label{eq:heff-2}
\end{eqnarray}

As the critical point approaches, this procedure can be applied again and again. In other words, from the SP equation for the first generation block variable ${\bf m}^I$, we can get the second generation blocks, and second generation block variables ${\bf m}^{II}$. Then from the second generation we get the third generation variables ${\bf m}^{III}$ , etc.  Then, we should obtain infinite generations of  blocks at the critical point. However, for the real systems, the number of generations wil not be so large. For example, Let the lattice spacing, $a=1{\AA}=10^{-10}m$, the first generation block sizes are generally $a_I\sim 10^{2}-10^{3}{\AA}$ \cite{wu2} if we set $\Delta \sim 0.3$. Accordingly, the second generation block sizes are $a_{II} \sim 10^{2}-10^{3} a_I$ assuming the variance of $r_I$ also being $\Delta_I \sim 0.3$. Then for a macroscopic sample with size $10^{-2}m$, there will be $3$ or $4$ generations. This recursive structure is like a Chinese (nested) boxes.

\section{The Chinese boxes structure in phase space}

\begin{figure}
 \begin{center}
    \resizebox{8cm}{7cm}{\includegraphics{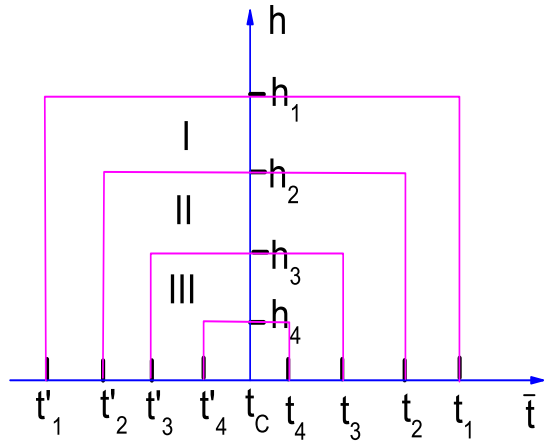}}
  \end{center}
\vskip -1cm

\caption{The schematic of the regime in which the ground state of each generation blocks dominate.}
\end{figure}

As shown above, for this model, there should be no simple asymptotic scaling behavior because there appear new generation blocks with larger size endlessly as we approach the critical point.

Because there are excited state SP  solutions, it seems not able to apply the RG to this model directly. Here we propose an approximate way to investigate the scaling behavior in this model. We divide the phase space into parts as shown in figure 1.  This figure is only schematic and the scale is not linear to the real values.

Let us to see the outside of the box. For $h=0$ and $\bar{t}>t_1 \sim \Delta^{4/(4-d)}$, there is no block and the SP point solution is $\b{\bf m}=0$. In this case, The system can be described by the LGW Hamiltonian with $\Phi_0=0$, and the RG can be applied with the help of replica trick to average the disorder. This has been done in the conventional RG theories 40 years ago \cite{harrisa,khm,grinstein}. For $h=0$ and $\bar{t}<t'_1$, the interactions between the first generation blocks are so strong£¬say the average $<J^I_{ij}>$ larger then 1, that the magnetization saturates and only the ground state SP solutions accounts. The order of $t_1-t'_1$ is $\sim \Delta^{8/3}$ for 3-dimensional system. It is very small.

For $\bar{t}=t_c$, where $t_c$ is exact critical point, there should be infinite generations of block if there is no field. As the field is applied, for $h>h_1$ the first generation superspins are totally polarized according to the discussion after Eq. (\ref{eq:feff}) where the effective field on the superspins $h_i^I \sim 10$. Then only the ground state SP solution for the first generation blocks accounts. The higher generations will be absent. In three dimension, the estimated value of $h_1\sim 10^{-3}$ assuming $\Delta=0.3$.

Now let us see the first layer. For $h<h_1$ and $t_2<\bar{t}<t_1$, there appear first generation blocks and their interactions are weak so that there is no second generation block. There is only ground state SP solution for the first generation superspins. Then the system can be described by the Hamiltonian Eq. (\ref{eq:hI}) with the ground SP solution. This Hamiltonian can be dealt with RG and replica trick as the usual LGW with zero SP solution. For $h_2<h<h_1$ and $t'_2<\bar{t}<t_2$, there are second generation blocks but there are totally polarized, where the effective fields on the second generation superspins are strong, say $h^{II}_i \sim 10$. Then the system is described by Hamiltonian Eq. (\ref{eq:heff-2}) with all second generation supespins $s^{II}_i$ are parallel to the external field.

We can similarly discuss the second, third layer, fourth layer, etc.  In the out layer, the system is described by the usual LGW Hamiltonian with the ground state SP solution and no blocks. In the first layer, the system is described by the Hamiltonian with ground state SP solution for the first generation superspin. In the second layer,  the system is described by the Hamiltonian with ground state SP solution for the second generation superspin. And so on.

The non-perturbative effect has been taken into account in these effective Hamiltonians with block superspins, so the RG can be applied to them directly. The effect of disorder, such as the spatial fluctuation of reduced temperature $t^I, t^{II},\cdots$ and the effective field ${\bf h}^I,{\bf h}^{II},\cdots$ can be dealt with the replica trick \cite{khm,grinstein}.

The boxes in figure 1 draw approximate regimes, where different generation of blocks dominate. However there is no clear-cut dividing lines between these regimes. For example, for $\bar{t}>t_1=\Delta^{4/(4-d)}$, the first generation block density is very small, but it is not zero.

No matter how are the scaling behaviors of these Hamiltonian of different generation, the system can not be described by a single scaling function, which is successful for the pure system. In different regime, it should be described by different scaling functions. Outside the box, there is no blocks, the system is described by the usual RG theories obtained 40 years ago \cite{khm,grinstein}. In the first layer with the presence of first generation block, there appear a new length scale $a_I$. The scaling function must take it into account. The scaling indexes, such as $\nu,\beta$, may be different from those for the Hamiltonian without the new terms. Nonetheless the scaling function should be different. The higher the generation  is, the more the new length scales there are. Therefore the scaling functions should be different in different regimes.

\section{Summary and discussion}

We propose a new possible way to understand the critical phenomena in the disordered systems. This model can be incorporating with the existing theories and can explain the experiments qualitatively.

As mentioned in the last section, the conventional RG theoy \cite{khm, grinstein} can describe the critical behavior in the outer layer in figure 1.
For the present Monte Carlo simulation, the usual size of lattice is about $10^{2}-10^{3}$. This is the order of the usual block size. The lattices are not enough to contain many blocks, therefore this kind finite size systems are described by the Hamiltonian given by Eq. (\ref{eq:h0}). One can apply replica trick and RG to it directly. This is why most of the Monte carlo simulation support the Harris' Criterion and the conventional RG theory.

The nonzero SP solutions above the critical temperature have been discussed in the RG with replica symmetry breaking \cite{dotsenko1995} and the rare regions theory \cite{vojta-jpa}. In those theories, the locally ordered regions are independent from each other. On the contrary, in the block model, the coupling between the blocks dominates the phase transitions. This is the most important difference of this discussion with the previous theories.

If the disorder is not so strong, the regime of anomalous critical behavior due to the blocks is very small and can not be accessible in the usual experiments and Monte Carlo simulation. However the experiments on the Griffiths phase, especially the experiments for the doped MnO materials indeed show anomalous properties\cite{salamon,salamon1,triki}, which can not be understood in the conventional framework of theory.

Chan, Goldenfeld and Salamon obtained new scaling relations from a heuristic calculation of the leading order essential singularity in the distribution of Yang-Lee zeroes. \cite{salamon1}. They give a physical picture for this derivation. A disordered ferromagnet can be thought of as an ensemble of weakly interacting, finite-sized ferromagnetic clusters. The probability of the cluster with $L$ follows Poisson's distribution $exp(-cL^d)$. The picture of block model is different. The sizes of each generation blocks are assumed to follow Gaussian's distribution. The average size of each generation blocks is a new emergent characterizing  length scale.

The anomalous critical behaviors in the  experiment for the doped MnO materials \cite{salamon,salamon1,triki} can be explained in the block model qualitatively.

An remarkable effect in the Griffiths phase is the sharp downturn or knee in the inverse susceptibility \cite{salamon}. As the temperature decreases, the inverse susceptibility deviate from a straight line and drop drastically. This effect can be explained with the appearance of LORs. There is a peak in the local susceptibility just when a LOR appears, and the temperatures  of the peak for different LORs are different. At the high temperature side, there is no LOR, the inverse susceptibility is a straight line. As the temperature lowers, some LORs appears, the local susceptibility develop peaks for these LORs, then they cause a drop in the inverse susceptibility.

For $h>h_1$, there is no excited state for the first generation block. Only the ground state SP solution accounts. For $h>h_2$, there is no excited state for the second generation block. The Hamiltonian ends at the first generation blocks, etc. Therefore, for different field range, the critical behavior obeys different scaling  functions. This seems to be able to explain the heat capacity results in Fig. (6.a) in reference \cite{salamon}. The scaling behaviors for $0.5T,1T$ and $5T,7T$ obey different scaling functions. We can assume $0.5T,1T$  are less than the threshold for certain generation blocks and $5T,7T$ are larger, so the critical behaviors obey different scaling functions.

Another anomalous behavior in the Griffiths phase is the unusually large value of the critical index $\delta$, which varies from $5.1$ to $16.9$ for different materials, if the critical behavior is assumed to obey the scaling laws. This is inconsistent with any known  universality class, but indicative of a very rapid and dramatic rise in magnetization. This feature can be understood with the block model qualitatively. The blocks are already locally ordered. The zero magnetization is due to the fluctuation of block superspins. As discussed above, very small external field can induced great effective field on the superspins to polarize them. This can explain the rapid and dramatic rise in magnetization. In addition, the stronger the disorder is, the deeper the blocks are ordered and the larger range of phase space is where the blocks play important role. Therefore for weak disorder, the critical behavior is the same as that without the blocks if the system is not close the critical point enough. For strong disorder, the anomalous critical behavior due to the blocks can be shown at not very close to the critical point.

Of course, our proposal should be judged by the experiments. If this model is valid, the emergent length scale of the blocks should be discovered. Some experiments indeed show the existence of the blocks. For example, two length scales of magnetic correlations are observed in the Invar $Fe_{75}Ni_{25}$ alloy above $T_C$ by means of small-angle neutron scattering and neutron depolarization \cite{grig1,grig2,grig3}. One length scale is attributed to the critical fluctuation, another one is attributed to the ferromagnetic clusters in the paramagnetic phase. These clusters should be the first generation blocks. If the experiment is improved further and the second generation blocks is observed, the Chinese boxes structure is verified.

As we know, a popular theory for the Griffith phase is rare region theory. However the probability of rare region is exponentially small, their effects are usually invisible. So the rare region theory seems to not able to explain the anomalous observable effects in the Griffiths phase. On the contrary, the probability of locally order regions is finite, they can induce observable effects.

Here we give a remark on the significance of the block in the block model. As we know, in the real space RG theory, the first step is to divided the system into blocks. These two ``blocks" are different, the former one is self-organized and the size and shape are determined by the random temperature; while the latter one is determined manually and can be arbitrary size.

We have also studied excited state local mean field solutions for the Blume-Capel model with random bond \cite{wu2} and De-Gennes-Bogliubov equaiton for the strongly disordered Hubbard model with negative-U \cite{wu3}. As we know, these solutions are all mean filed like. They show the same character: the system is self-organized into blocks, which are coupled with each other like superspins. In other words, the picture of blocks is common in the phase transition in disordered systems.

This proposal is waiting for a lot of future works. The 3-dimensional SP solution has not been studied because it needs huge memory and computation time. The effect of the external field is being studied numerically. The renormalization group studies on the effective Hamiltonian with blocks are difficult, however they are valuable. The scaling taking the block sizes into account are also prospective.

\appendix

\section{}

For a local region with size much larger than a block, we can assume the order parameter is along the direction of ${\bf e}_1$. According to the setting in Eq. (\ref{eq:gold}), we can show that the Gaussian approximation for the transverse fluctuation around the ground state
\begin{equation}
\delta F_{G}=\frac{1}{a^d}\int d{\textbf x} {1 \over 2} a^2 \Phi_0^2|\bigtriangledown {\textbf n}|^2
\label{eq:gaussian-appro}
\end{equation}
is approximately equal to Eq. (\ref{eq:gaussian-appro1}). Let
\begin{equation}
\tilde{n}_i=\tilde{m}_i/\b{$m$}=\tilde{m}_i/\Phi_0,
\end{equation}
we have
\begin{equation}
\tilde{n}_1=\delta (\sqrt{1-\sum_{i=2}^{p}n_i^2})\approx -\frac{1}{2}\sum_{i=2}^p\tilde{n}_i^2
\end{equation}
therefore we can ignore the variation of the component $n_1$ in the first order approximation. Then we have
\begin{eqnarray}
\int d {\bf x} \Phi_0^2|\bigtriangledown {\bf n}|^2 &  \approx  & \int d {\bf x} \Phi_0^2 \sum_{i=2}^p|\bigtriangledown \tilde{n}_i|^2 \nonumber \\
& = & -\int d {\bf x} \tilde{n}_i\bigtriangledown \cdot (\Phi_0^2 \bigtriangledown \tilde{n}_i)
\end{eqnarray}
Using
\begin{equation}
\Phi_0^2 \bigtriangledown \tilde{n}_i= \Phi_0^2 \bigtriangledown (\tilde{m}_i/\b{$m$})=\Phi_0\bigtriangledown \tilde{m}_i-\tilde{m}_i\bigtriangledown \Phi_0,
\end{equation}
we get
\begin{equation}
\int d{\bf x}  \bigtriangledown(\Phi_0^2 \bigtriangledown \tilde{n}_i)= \int d{\bf x}  [\Phi_0\bigtriangledown^2 \tilde{m}_i-\tilde{m}_i \bigtriangledown^2 \Phi_0].
\end{equation}
Then we obtain
\begin{equation}
\int d {\bf x} \Phi_0^2|\bigtriangledown {\bf n}|^2=\int d{\bf x} [-\tilde{m}_i\bigtriangledown^2 \tilde{m}_i+\frac{\tilde{m}_i^2}{\Phi_0}\bigtriangledown^2 \Phi_0].
\end{equation}
Since $\Phi_0$ satisfies the Eq. (\ref{eq:spe1}), we prove the Eq.(\ref{eq:gaussian-appro1}) from Eq.(\ref{eq:gaussian-appro}).

\end{document}